\documentclass[11pt]{article}
\usepackage{amsmath,amsfonts,amssymb,graphicx,hyperref,xcolor,booktabs,array,float,enumerate}
\usepackage[margin=1in]{geometry}
\usepackage{mathtools}
\usepackage{tensor}    
\usepackage[most]{tcolorbox}

\newtcolorbox{contributionbox}{
  enhanced,
  colback=white,
  colframe=black,
  boxrule=0.8pt,
  arc=2mm,
  left=2mm,
  right=2mm,
  top=1mm,
  bottom=1mm,
  title=\bfseries Principal Contributions,
  fonttitle=\bfseries,
}

\hypersetup{hidelinks}
\newcommand{\rev}[1]{#1}
\newcommand{\Caputo}{\prescript{C}{}{D}_t^{\alpha}}
\title{\bf Impulsively Driven Fractional Relaxation:\\
Exact Response, Scaling Laws, Frequency-Domain Signatures, and Memory-Induced Crossover Structures}
\author{Koichi Nakagawa\footnote{Corresponding author: nakagawa@hoshi.ac.jp}\\Hoshi University, Tokyo, Japan}
\date{}

\begin{document}
\maketitle

\begin{abstract}
\rev{We investigate periodically impulsive fractional relaxation as a minimal model of nonlocal dissipative dynamics under repeated external stimulation. The free relaxation law is classical, but the driven problem introduces an additional timescale whose competition with the intrinsic fractional-memory scale generates nontrivial accumulation and crossover behavior. We derive the exact Laplace-domain and time-domain responses, establish the sparse-forcing scaling of the long-time averaged response, and obtain an analytical crossover interval proportional to the inverse fractional power of the relaxation coefficient.}

\rev{To separate established properties of Mittag--Leffler relaxation from the new effects of forcing, we explicitly compare the unforced, impulsive, and periodically driven settings. Extended long-time simulations verify the algebraic tail, while an independent L1 discretization provides a numerical benchmark for the exact solution. We further connect the relaxation kernel to measurable frequency-domain quantities through the Cole--Cole complex susceptibility and discuss interpretations in dielectric and viscoelastic relaxation, anomalous transport, non-Markovian open systems, and intermittent reinforcement. The resulting framework supplies an analytically solvable reference model for memory accumulation under repeated driving.
The present analytical framework therefore provides a bridge
between fractional relaxation theory and experimentally
accessible time- and frequency-domain observables.}
\end{abstract}

\section{Introduction}
Fractional relaxation is a standard mathematical representation of systems whose present evolution depends on a broad distribution of past times. It has been used in viscoelasticity, dielectric relaxation, anomalous diffusion, transport with memory, and non-Markovian dynamics \cite{Podlubny1999,Mainardi2010,Metzler2000,Hilfer2000,Tarasov2011,Gorenflo2014,Gorska2024,Barbero2024}. In the absence of forcing, the Caputo relaxation equation and its Mittag--Leffler solution are well established, including their stretched-exponential-like intermediate behavior and algebraic long-time tail \cite{Mainardi1996,Gorenflo2014}. These properties are therefore treated here as background rather than as new results.

\rev{A distinct question arises when a relaxing system is repeatedly perturbed by short external events. Periodic electric-field pulses, mechanical loading cycles, repeated injections, control actions, or reinforcement events introduce a forcing interval that competes with the intrinsic memory time. Although impulsive fractional differential equations have been investigated extensively from the viewpoints of existence, controllability, and stability \cite{Feckan2012,Stamova2015,RenWang2022,Kavitha2022,Mathiyalagan2023,Prabu2024}, considerably less attention has been paid to closed-form scaling relations that connect pulse spacing, fractional memory, and a measurable accumulated response.}

\rev{The scientific gap addressed here is thus not the solution of the free Caputo equation, nor the use of the Laplace transform itself. Unlike previous studies,
which mainly addressed
existence,
stability,
or controllability
of impulsive fractional equations,
the present work derives
closed-form scaling laws
and experimentally interpretable
crossover criteria. It is the explicit analytical organization of periodically forced fractional relaxation into a similarity variable, a sparse-forcing scaling law, and a crossover boundary that can be compared directly with parameter-space diagrams and frequency-domain response functions.}

The present work makes four contributions. First, it derives the exact response to a periodic delta-pulse train as a superposition of delayed two-parameter Mittag--Leffler kernels. Second, it identifies a dimensionless competition parameter between the forcing interval and the fractional-memory time. Third, it derives the asymptotic scaling of the long-time averaged response and the corresponding crossover interval. Fourth, it connects the time-domain model to complex susceptibility and provides numerical benchmarks and physically interpretable parameter mappings.

\rev{The restriction $0<\alpha\leq 1$ is deliberate. This range describes completely monotone relaxation for positive relaxation coefficient, whereas $1<\alpha<2$ introduces oscillatory or inertia-like behavior. The latter regime is important but belongs to fractional relaxation--oscillation rather than to the monotone memory-decay problem considered here \cite{Mainardi1996}.}

\begin{contributionbox}

The principal novelty of the present work is not the
Mittag--Leffler relaxation itself,
which is well established,
but the analytical characterization of periodically
impulsively driven fractional relaxation.

The main contributions are:

\begin{enumerate}[(i)]
\item An exact closed-form solution for arbitrary periodic impulse trains.
\item A universal crossover scaling law connecting impulse interval and fractional order.
\item Frequency-domain interpretation through complex susceptibility.
\item Analytical regime classification of sustained-retention and decay-dominated behavior.
\end{enumerate}

\end{contributionbox}

Recent developments have also focused on
generalized memory kernels
and stability properties
of fractional relaxation equations
\cite{CJP2024}.
Recent studies have considered
impulsive fractional equations
mainly from the viewpoints of
existence,
stability,
and controllability
\cite{Kattan2024}.

The paper is organized as follows. Section~2 defines the model and its dimensions. Section~3 establishes the free response and long-time behavior. Section~4 derives the exact periodically forced solution. Section~5 obtains the scaling laws and crossover criterion. Section~6 presents numerical tests and regime diagrams. Section~7 gives physical interpretations. Section~8 connects the model to dielectric susceptibility. Section~9 compares the present framework with representative existing models. Sections~10 and 11 contain the discussion and conclusions, and the appendices provide derivational and numerical details.

\section{Mathematical formulation and dimensional consistency}
Let $R(t)$ denote a scalar relaxing observable. Depending on the application, it may represent normalized polarization, stress, concentration excess, population imbalance, or retained response. For $0<\alpha<1$, the Caputo derivative is
\begin{equation}
\Caputo R(t)=\frac{1}{\Gamma(1-\alpha)}\int_0^t \frac{R'(u)}{(t-u)^\alpha}\,du.
\label{eq:caputo}
\end{equation}
The limiting case $\alpha=1$ recovers the ordinary derivative.

\rev{The one- and two-parameter Mittag--Leffler functions are defined, respectively, by}
\begin{align}
E_{\alpha}(z)&=\sum_{n=0}^{\infty}\frac{z^n}{\Gamma(\alpha n+1)},\label{eq:ml1}\\
E_{\alpha,\beta}(z)&=\sum_{n=0}^{\infty}\frac{z^n}{\Gamma(\alpha n+\beta)}.\label{eq:ml2}
\end{align}
These series are entire for positive $\alpha$ \cite{Gorenflo2014}.

The free relaxation equation is
\begin{equation}
\Caputo R(t)=-\lambda R(t),\qquad R(0)=R_0.
\label{eq:free}
\end{equation}
Because the Caputo derivative has dimension $T^{-\alpha}$, dimensional consistency requires
\begin{equation}
[\lambda]=T^{-\alpha}.
\label{eq:lambda_dim}
\end{equation}
The argument $\lambda t^\alpha$ is therefore dimensionless. For the impulsively driven model below, $\rho$ has the dimension of $R$ when the delta pulses are interpreted as instantaneous state increments.

\rev{For $R_0\geq0$, $\lambda>0$, and $0<\alpha\leq1$, the solution is nonnegative and completely monotone. This property is central to the physical interpretation as relaxation without overshoot. Extending the analysis to $1<\alpha<2$ would require an additional initial datum and would describe fractional relaxation--oscillation, which is outside the present scope.}

\section{Free relaxation and long-time asymptotics}
Taking the Laplace transform of Eq.~\eqref{eq:free} gives
\begin{equation}
s^\alpha \widetilde R(s)-s^{\alpha-1}R_0=-\lambda\widetilde R(s),
\end{equation}
so that
\begin{equation}
\widetilde R(s)=\frac{s^{\alpha-1}R_0}{s^\alpha+\lambda},\qquad
R(t)=R_0E_\alpha(-\lambda t^\alpha).
\label{eq:free_solution}
\end{equation}
For $0<\alpha<1$,
\begin{equation}
E_\alpha(-\lambda t^\alpha)\sim \frac{1}{\lambda\Gamma(1-\alpha)}t^{-\alpha},\qquad t\to\infty.
\label{eq:free_asymptotic}
\end{equation}

\begin{figure}[H]
\centering
\includegraphics[width=0.98\textwidth]{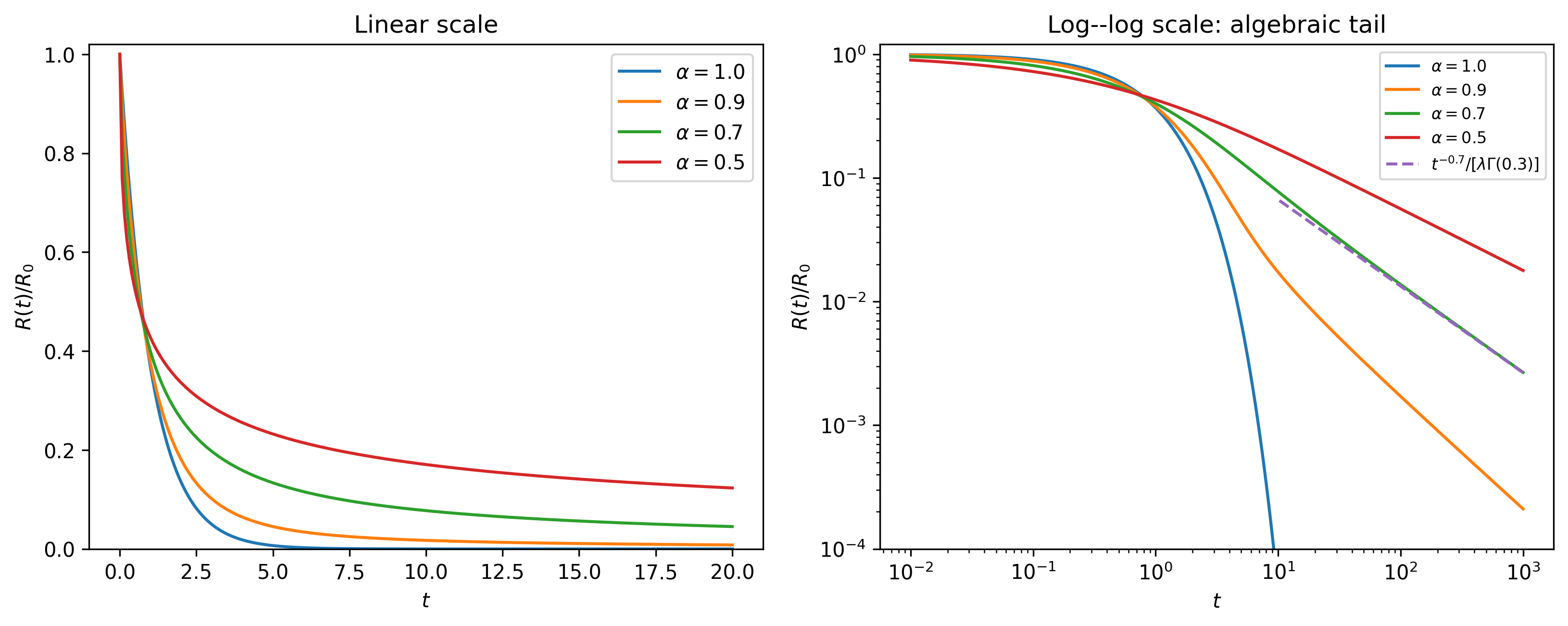}
\caption{Extended free-relaxation benchmark for $\lambda=1$. Left: linear scale. Right: log--log scale, including the asymptotic guide for $\alpha=0.7$. The algebraic tail is visible only when the time window is extended well beyond the initial transient.}
\label{fig:longtime}
\end{figure}

\rev{Figure~\ref{fig:longtime} extends the time window by several decades. The linear panel shows the separation from exponential relaxation, while the log--log panel directly displays the algebraic tail. The dashed guide is the known asymptotic form and is included as a benchmark, not as a claimed novelty.}

\begin{figure}[H]
\centering
\includegraphics[width=0.70\textwidth]{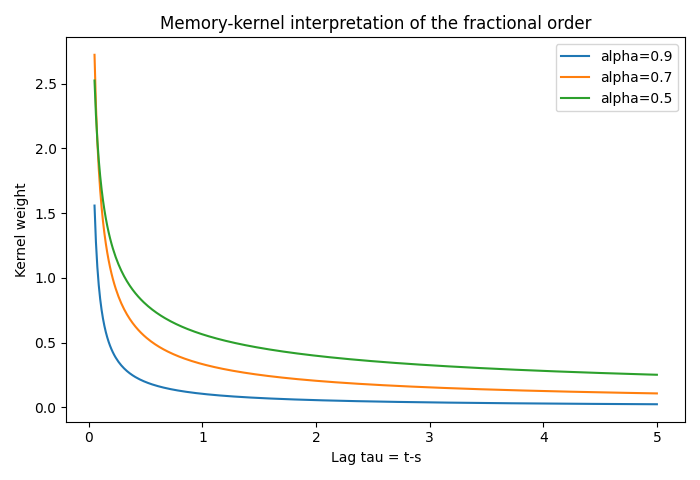}
\caption{Power-law memory kernels for representative fractional orders. Smaller $\alpha$ corresponds to a longer-tailed memory contribution.}
\label{fig:kernel}
\end{figure}

The memory-kernel interpretation follows directly from Eq.~\eqref{eq:caputo}. Smaller $\alpha$ produces a more persistent weighting of the past, as illustrated in Fig.~\ref{fig:kernel}.

\section{Exact response to periodic impulsive forcing}
We consider
\begin{equation}
\Caputo R(t)=-\lambda R(t)+\rho\sum_{k=1}^{\infty}\delta(t-k\Delta t),
\qquad R(0)=R_0,
\label{eq:driven}
\end{equation}
where $\rho$ is the pulse increment and $\Delta t$ is the pulse interval. The Laplace transform is
\begin{equation}
\widetilde R(s)=\frac{s^{\alpha-1}R_0}{s^\alpha+\lambda}
+\frac{\rho}{s^\alpha+\lambda}\frac{e^{-s\Delta t}}{1-e^{-s\Delta t}}.
\label{eq:laplace_driven}
\end{equation}
Inverting term by term yields
\begin{equation}
R(t)=R_0E_\alpha(-\lambda t^\alpha)
+\rho\sum_{k\Delta t<t}(t-k\Delta t)^{\alpha-1}
E_{\alpha,\alpha}\!\left[-\lambda(t-k\Delta t)^\alpha\right].
\label{eq:time_driven}
\end{equation}
Equation~\eqref{eq:time_driven} shows that each pulse produces a delayed memory trace. Unlike a reset, the impulse does not erase the earlier history; the response is an accumulated convolution of the pulse train with the fractional impulse kernel.

\begin{figure}[H]
\centering
\includegraphics[width=0.82\textwidth]{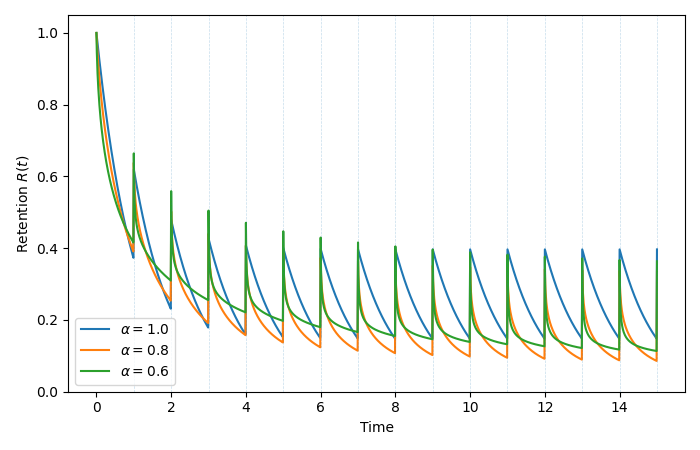}
\caption{Representative impulsively driven trajectories under identical forcing conditions. Fractional memory retains contributions from earlier pulses, producing stronger accumulation than the Markovian case.}
\label{fig:impulse}
\end{figure}

\rev{This distinction is the first genuinely forcing-induced feature of the model. The classical Mittag--Leffler tail belongs to the free equation; the nonlocal superposition of repeated pulse traces and its dependence on $\Delta t$ belong to the driven problem.}

\section{Scaling law and crossover criterion}
Define the long-time averaged response
\begin{equation}
\overline R=\frac{1}{T}\int_{T_0}^{T_0+T}R(t)\,dt,
\label{eq:Rbar}
\end{equation}
with $T_0$ chosen after the initial transient. The intrinsic memory time is obtained from
\begin{equation}
\lambda t_m^\alpha=O(1),\qquad t_m\sim\lambda^{-1/\alpha}.
\label{eq:tm}
\end{equation}
The dimensionless forcing ratio
\begin{equation}
\Theta=\frac{\Delta t}{t_m}=\Delta t\,\lambda^{1/\alpha}
\label{eq:theta}
\end{equation}
organizes the dynamics.

For sparse forcing, the surviving fraction of a pulse after one interval obeys
\begin{equation}
E_\alpha(-\lambda\Delta t^\alpha)
\sim\frac{\Delta t^{-\alpha}}{\lambda\Gamma(1-\alpha)}.
\end{equation}
Hence
\begin{equation}
\overline R\sim C(\alpha,\lambda,\rho)\,\Delta t^{-\alpha},
\qquad \Theta\gg1.
\label{eq:Rbar_scaling}
\end{equation}
A crossover interval may be defined by a fixed surviving fraction $0<\eta<1$,
\begin{equation}
E_\alpha(-\lambda\Delta t_c^\alpha)=\eta.
\end{equation}
If $E_\alpha(-z_\eta)=\eta$, then
\begin{equation}
\Delta t_c=\left(\frac{z_\eta}{\lambda}\right)^{1/\alpha}
\propto\lambda^{-1/\alpha}.
\label{eq:crossover}
\end{equation}

\rev{Equations~\eqref{eq:Rbar_scaling} and \eqref{eq:crossover} are the central analytical results. They do not assert a thermodynamic phase transition. Rather, they quantify a smooth change between pulse-overlap and pulse-isolation regimes.}

\section{Numerical verification and parameter-space structure}
\begin{figure}[H]
\centering
\includegraphics[width=0.78\textwidth]{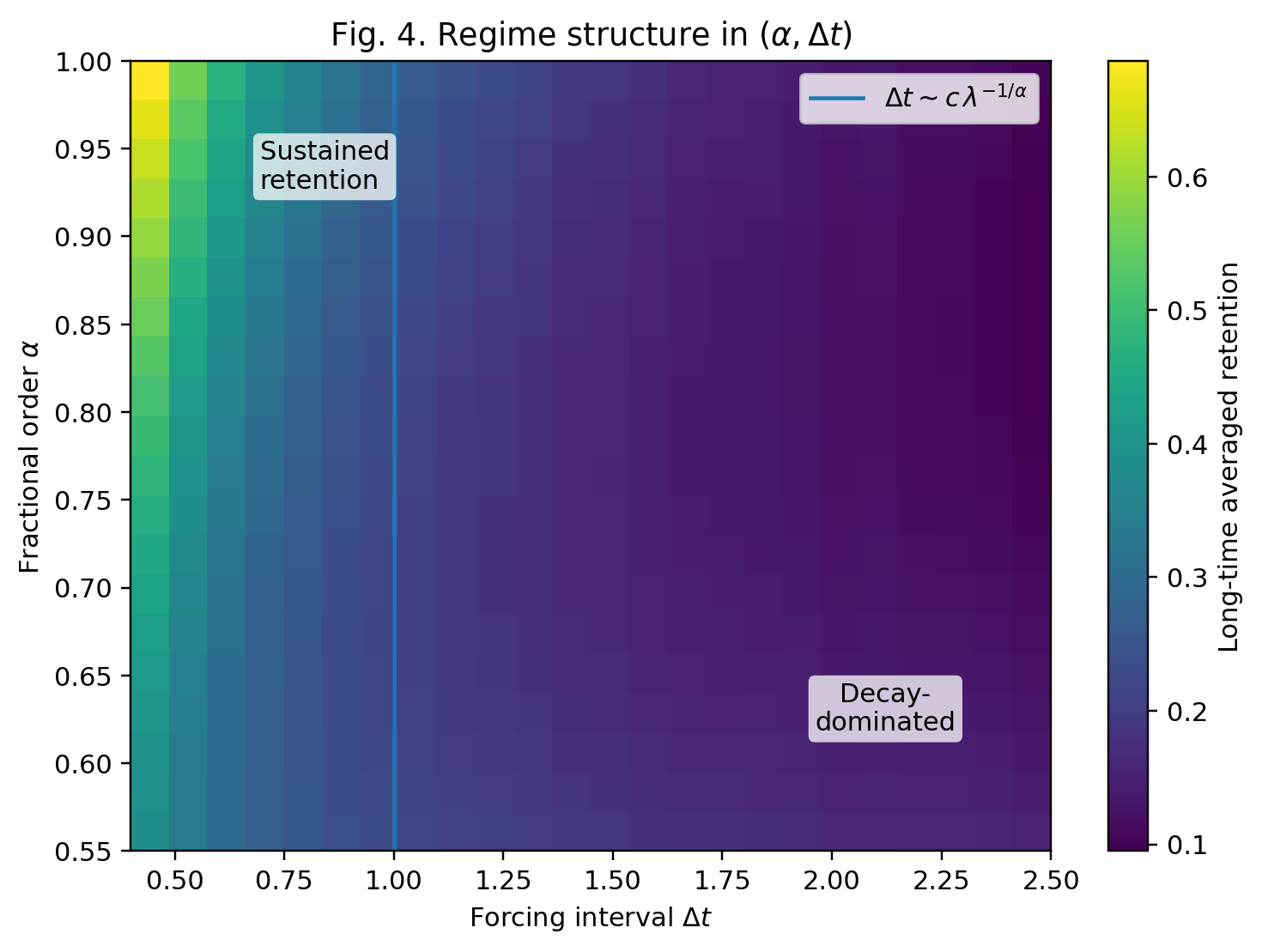}
\caption{Long-time averaged response in the $(\alpha,\Delta t)$ plane. The overlaid guideline represents the scaling $\displaystyle\Delta t_c=\left(\frac{z_\eta}{\lambda}\right)^{1/\alpha}$.}
\label{fig:phasealpha}
\end{figure}

\begin{figure}[H]
\centering
\includegraphics[width=0.78\textwidth]{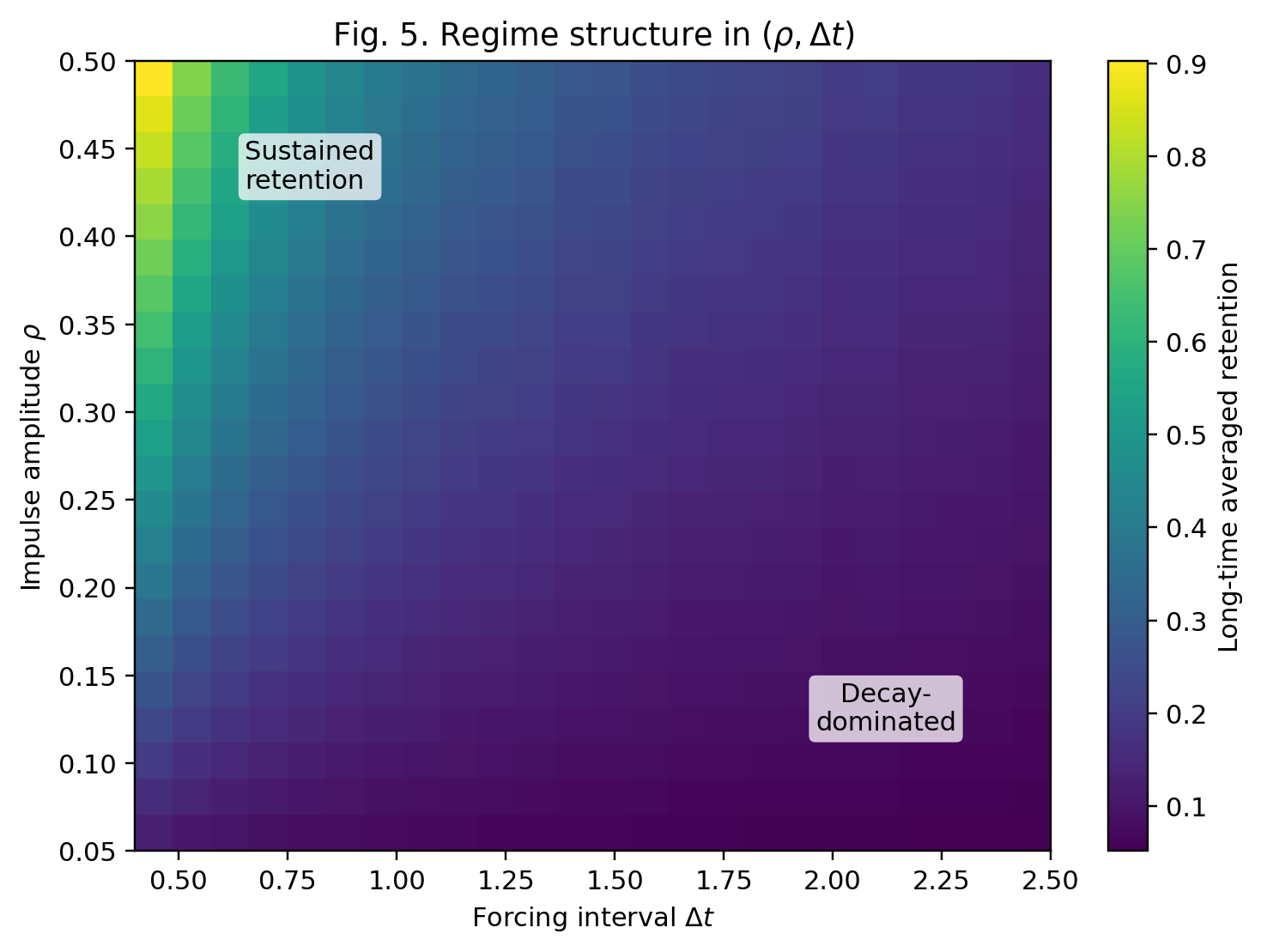}
\caption{Long-time averaged response in the $(\rho,\Delta t)$ plane. Increasing pulse amplitude and decreasing pulse spacing favor accumulated response.}
\label{fig:phaserho}
\end{figure}

The analytical relations are compared with direct numerical calculations. Figure~\ref{fig:phasealpha} shows the averaged response in the $(\alpha,\Delta t)$ plane, with the analytical crossover guideline superimposed. Figure~\ref{fig:phaserho} shows the corresponding dependence on pulse amplitude.

\begin{figure}[H]
\centering
\includegraphics[width=0.98\textwidth]{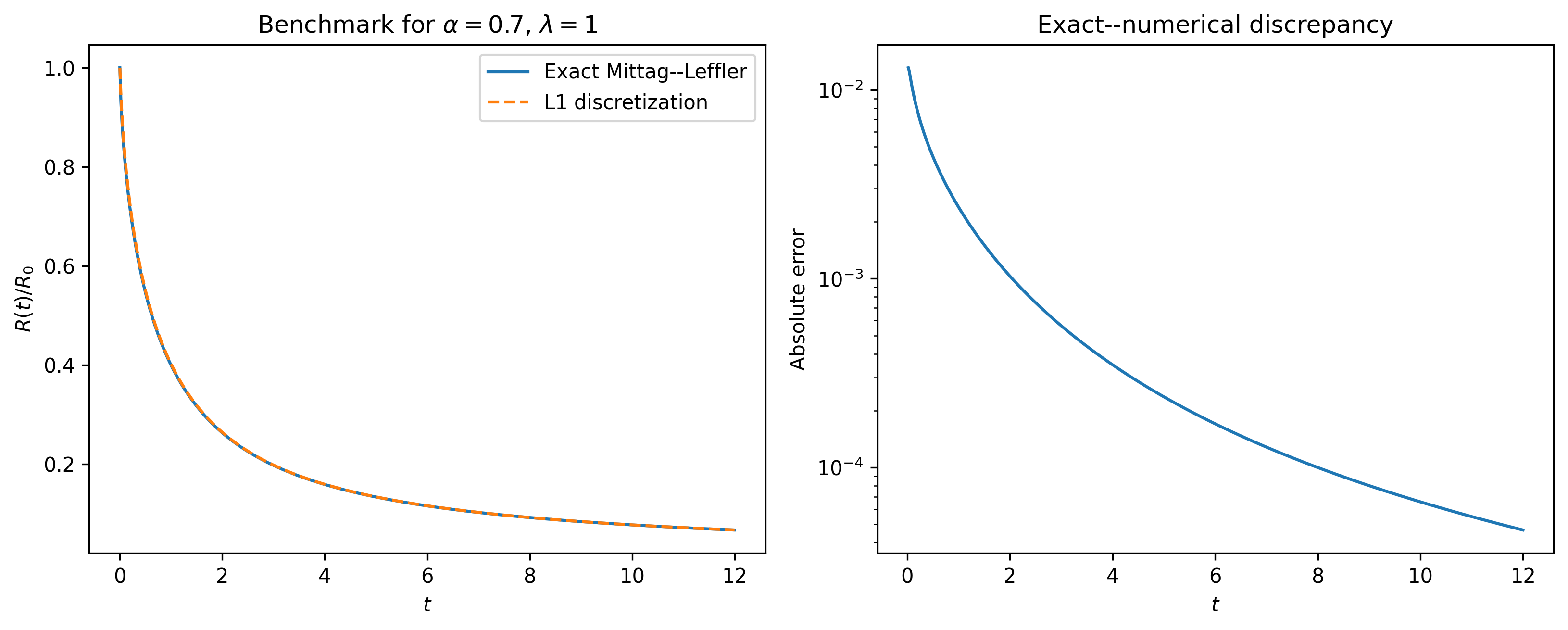}
\caption{Independent numerical benchmark for $\alpha=0.7$ and $\lambda=1$. Left: exact Mittag--Leffler solution and L1 discretization. Right: absolute discrepancy.}
\label{fig:benchmark}
\end{figure}

\rev{To provide an independent numerical benchmark, Eq.~\eqref{eq:free} was also solved by the L1 approximation of the Caputo derivative. Figure~\ref{fig:benchmark} compares the L1 solution with the exact Mittag--Leffler response. Their close agreement verifies the implementation used in the numerical calculations.}

\section{Physical interpretation of the parameters}
\rev{Although the model is intentionally minimal, its parameters possess direct interpretations across several applications. Table~\ref{tab:parameters} summarizes representative mappings.}
\begin{table}[H]
\centering
\caption{Representative physical interpretations of the model variables.}
\label{tab:parameters}
\begin{tabular}{p{0.12\textwidth}p{0.25\textwidth}p{0.50\textwidth}}
\toprule
Symbol & Mathematical role & Representative physical meaning\\
\midrule
$R(t)$ & Relaxing state & Polarization, stress, concentration excess, population imbalance, retained response\\
$\alpha$ & Memory exponent & Breadth and persistence of the relaxation-time distribution\\
$\lambda$ & Fractional relaxation coefficient & Intrinsic loss, damping, depolarization, forgetting, or leakage strength\\
$\rho$ & Pulse increment & Electric-field kick, stress increment, particle injection, control action, rehearsal strength\\
$\Delta t$ & Pulse interval & Driving period, loading interval, injection spacing, or rehearsal spacing\\
\bottomrule
\end{tabular}
\end{table}

In dielectric systems, $R$ may be normalized polarization and the pulses short field excitations. In viscoelastic systems, $R$ may be stress or strain memory and the pulses loading increments. In transport models, $R$ may be a coarse-grained excess population and the pulses injections. In reinforcement-like processes, $R$ is retained activation, $\rho$ is rehearsal strength, and $\Delta t$ is the spacing interval.

\rev{The crossover condition has a transparent physical meaning. When $\Delta t\ll t_m$, successive pulses overlap before the preceding trace has decayed; when $\Delta t\gg t_m$, each trace is nearly isolated. Thus increasing $\lambda$ shortens the memory window, while decreasing $\alpha$ broadens the distribution of relevant past times.}

\section{Connection with dielectric relaxation and complex susceptibility}

To connect the time-domain response with experimentally measurable dielectric quantities, we consider the complex susceptibility. For a linear relaxation function $\phi(t)$ normalized by $\phi(0)=1$, the complex susceptibility may be written as
\begin{equation}
\chi^*(\omega)=\chi_\infty+\Delta\chi\left[1-i\omega\int_0^\infty e^{-i\omega t}\phi(t)\,dt\right].
\label{eq:sus_definition}
\end{equation}
For Mittag--Leffler relaxation with characteristic time $\tau=\lambda^{-1/\alpha}$, this yields the Cole--Cole form
\begin{equation}
\chi^*(\omega)=\chi_\infty+
\frac{\Delta\chi}{1+(i\omega\tau)^\alpha}.
\label{eq:colecole}
\end{equation}
The Debye result is recovered at $\alpha=1$ \cite{ColeCole1941}. Broader empirical dielectric responses are often described by the Havriliak--Negami generalization \cite{HavriliakNegami1967}.

\begin{figure}[H]
\centering
\includegraphics[width=0.98\textwidth]{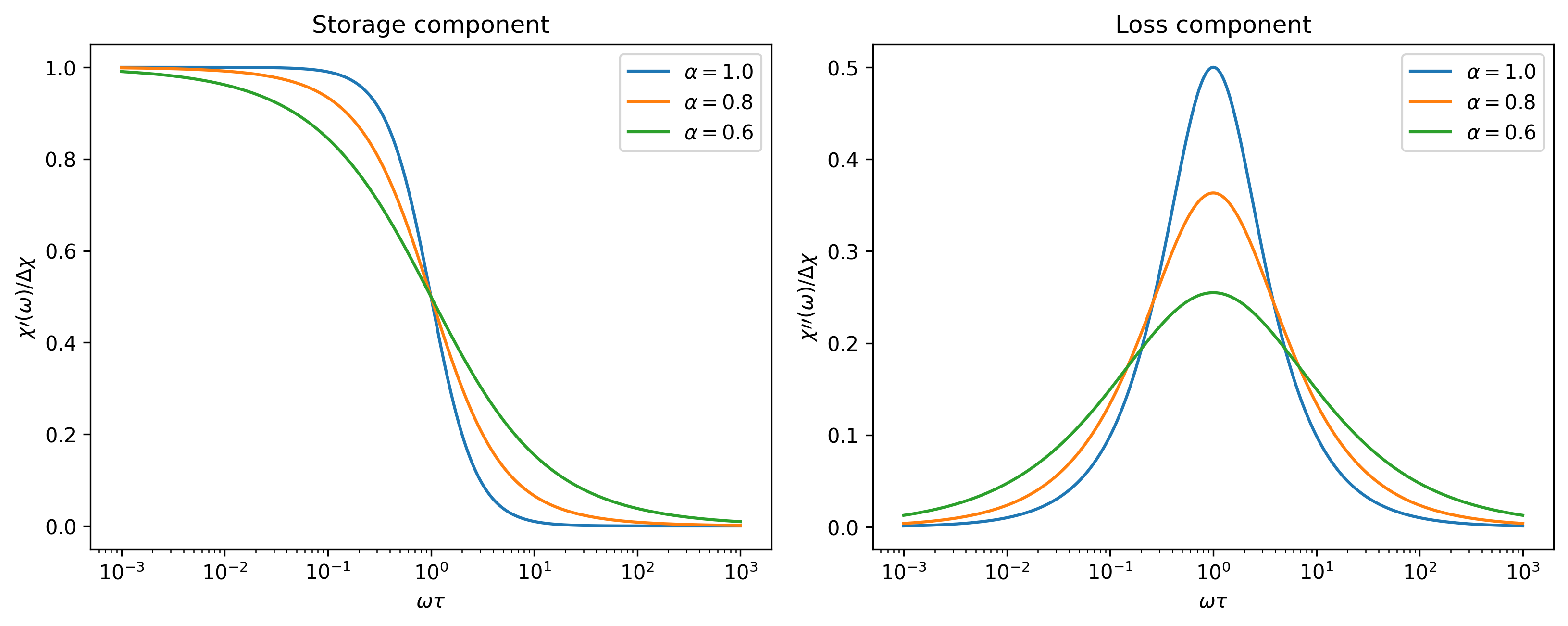}
\caption{Cole--Cole complex susceptibility associated with fractional relaxation. Left: normalized storage component. Right: normalized loss component. Smaller $\alpha$ broadens the response. }
\label{fig:susceptibility}
\end{figure}

\rev{Figure~\ref{fig:susceptibility} shows the storage and loss components of Eq.~\eqref{eq:colecole}. Decreasing $\alpha$ broadens the loss peak and spreads the response over a wider frequency range. This calculation provides a concrete measurable signature of the same memory exponent that controls the time-domain decay and the pulse-spacing crossover.}

Under periodic pulsing, the output spectrum is the product of the pulse-train spectrum and the fractional transfer function $(s^\alpha+\lambda)^{-1}$ evaluated on the imaginary axis. The same formalism therefore predicts both time-domain accumulation and frequency-domain filtering. A detailed fit to a specific material is beyond the scope of this minimal study, but Eq.~\eqref{eq:colecole} establishes a direct route to dielectric spectroscopy and related response measurements. Recent fractional models of polymer dielectric relaxation illustrate the relevance of such connections \cite{Renteria2025}.

This suggests that the crossover identified in the
time domain should also appear
as a characteristic broadening
in the frequency domain.
Such frequency-domain signatures
may be directly compared with
broadband dielectric spectroscopy
or dynamic mechanical analysis,
providing a practical route
for experimental verification.

Recent reviews also emphasize that
fractional constitutive equations naturally describe
electrical impedance
and dielectric spectroscopy
in complex media
\cite{Barbero2024}.

\section{Comparison with representative existing models}
\rev{Table~\ref{tab:comparison} separates the established ingredients from the contribution of the present work. Free Caputo relaxation supplies the memory kernel and Mittag--Leffler decay. The impulsive-FDE literature supplies existence, stability, and controllability frameworks. The present study instead focuses on the exact periodically repeated response, its accumulated long-time observable, and analytical pulse-spacing crossover laws.}
\begin{table}[H]
\centering
\caption{Schematic comparison of representative model classes.}
\label{tab:comparison}
\scalebox{.9}{
\begin{tabular}{p{0.28\textwidth}cccc}
\toprule
Model class & Long memory & Repeated pulses & Closed crossover scaling & Regime map\\
\midrule
Ordinary Debye relaxation & No & No & No & No\\
Free Caputo/Mittag--Leffler relaxation & Yes & No & No & No\\
General impulsive fractional equations & Yes & Yes & Usually not the focus & Usually no\\
Present periodically driven model & Yes & Yes & Yes & Yes\\
\bottomrule
\end{tabular}
}
\end{table}

The comparison also clarifies the scope of the novelty claim. The Mittag--Leffler function, Laplace inversion, and free asymptotics are standard. The contribution is their organization into a periodically driven response with the experimentally interpretable competition parameter $\Theta$, the asymptotic averaged-response law, the crossover interval, and the frequency-domain connection.

\section{Discussion}
The exact solution shows that repeated impulses are filtered by a nonlocal response kernel rather than by a single exponential time constant. Consequently, the state immediately before a pulse depends on the full pulse history. The crossover is therefore an organization of timescales generated by the competition between $\Delta t$ and $t_m$, not a singular phase transition.

The sparse-forcing scaling provides a testable prediction: at sufficiently large spacing, the retained contribution of a previous pulse decreases algebraically with exponent $\alpha$. A time-domain experiment can vary the repetition interval and measure the residual response immediately before each pulse. Independently, a frequency-domain experiment can estimate $\alpha$ from the breadth of the susceptibility loss peak. Agreement between these two estimates would test the proposed link between memory and pulse-spacing response.

The model is also distinguishable from stochastic resetting. A reset replaces the state or its distribution by a prescribed state, whereas the present impulse adds a contribution without erasing the accumulated memory. The dynamics are therefore closer to memory-preserving reinforcement than to renewal by complete reset.

Several limitations should be emphasized. The pulse times and amplitudes are deterministic and identical; random spacing, distributed amplitudes, finite-width pulses, and state-dependent impulses may alter the prefactor or generate additional scales. The model is scalar and linear. Nonlinear constitutive laws, spatial transport, and operator-valued quantum dynamics require separate analysis. Finally, $1<\alpha<2$ would describe an oscillatory response and may produce resonance-like behavior under periodic forcing; this extension is left for future work.

Recent studies have further generalized
fractional relaxation equations
to include more general memory kernels
and stability analyses
\cite{CJP2024}.
The present work is complementary,
focusing instead on
periodically impulsive forcing
and analytically tractable
crossover structures.

Despite these limitations, analytical tractability is valuable. The model supplies a reference problem against which more complicated fractional systems can be compared. It also identifies which observations are inherited from free Mittag--Leffler relaxation and which arise specifically from repeated forcing.

The analytical framework developed here
may also be applicable to 
spaced learning,
repeated drug delivery,
pulse-controlled viscoelastic systems,
and intermittent control.

\section{Conclusions}
We studied a periodically impulsive Caputo relaxation equation in the completely monotone range $0<\alpha\leq1$. The free Mittag--Leffler response and its algebraic tail are established properties and were used as benchmarks. The new focus was the periodically driven problem.

The exact response is a superposition of delayed two-parameter Mittag--Leffler kernels. From this representation, the forcing interval and fractional-memory time combine into a dimensionless competition parameter. The sparse-forcing average obeys an algebraic spacing law, while a fixed-survival criterion yields a crossover interval proportional to $\lambda^{-1/\alpha}$. Numerical regime diagrams and an independent L1 calculation support the analytical construction.

The same memory exponent enters the Cole--Cole complex susceptibility, furnishing a direct connection to dielectric and viscoelastic response. The framework therefore links time-domain pulse accumulation, long-time scaling, and frequency-domain broadening within one minimal model.

Future extensions include random or finite-width pulses, nonlinear relaxation, distributed-order kernels, spatially extended systems, and the oscillatory range $1<\alpha<2$. The present model may serve as a benchmark model for those developments.

\section*{Declaration of Competing Interest}
The author declares no competing interests.

\section*{Data and code availability}
The data shown in the figures are generated numerically from the equations stated in the manuscript. The figure-generation script is supplied with the revision files.

\appendix
\section{Laplace inversion and impulse kernel}
Using
\begin{equation}
\mathcal L\{t^{\beta-1}E_{\alpha,\beta}(-\lambda t^\alpha)\}(s)
=\frac{s^{\alpha-\beta}}{s^\alpha+\lambda},
\end{equation}
with $\beta=1$ gives the free solution, while $\beta=\alpha$ gives
\begin{equation}
h_\alpha(t)=t^{\alpha-1}E_{\alpha,\alpha}(-\lambda t^\alpha),
\end{equation}
which is the impulse response. The periodic pulse train has transform
\begin{equation}
\sum_{k=1}^\infty e^{-sk\Delta t}=\frac{e^{-s\Delta t}}{1-e^{-s\Delta t}},
\end{equation}
leading directly to Eq.~\eqref{eq:time_driven} by the shift theorem.

\section{L1 numerical discretization}
For a uniform grid $t_n=n\Delta$, the L1 approximation is
\begin{equation}
\Caputo R(t_n)\approx \frac{1}{\Gamma(2-\alpha)\Delta^\alpha}
\sum_{j=0}^{n-1}b_j\left(R_{n-j}-R_{n-j-1}\right),
\end{equation}
where
\begin{equation}
b_j=(j+1)^{1-\alpha}-j^{1-\alpha}.
\end{equation}
Substitution into the free equation yields an explicit recursion for $R_n$. This method was used only as an independent benchmark; the analytical plots were generated from the exact Mittag--Leffler representation.

\end{document}